\documentclass[aps,twocolumn,groupedaddress,showpacs]{revtex4}
\usepackage{graphicx}%
\usepackage{latexsym}%
\begin{document}
\title{Nonresonant entrainment of detuned oscillators
 induced by common external noise}
\author{${}^1$Kazuyuki Yoshimura, ${}^1$Peter Davis, and ${}^2$Atsushi Uchida} 
\affiliation{
${}^1$NTT Communication Science Laboratories, NTT Corporation\\
 2-4, Hikaridai, Seika-cho, Soraku-gun, Kyoto 619-0237, Japan\\
${}^2$ Department of Electronics and Computer Systems, Takushoku University\\
 815-1 Tatemachi, Hachioji, Tokyo 193-0985, Japan
}
\date{\today}
\begin{abstract}

   We have found that
 a novel type of entrainment occurs
 in two nonidentical limit cycle oscillators
 subjected to a common external white Gaussian noise.
 This entrainment is anomalous in the sense that
 the two oscillators have different mean frequencies,
 where the difference is constant
 as the noise intensity increases,
 but their phases come to be locked for almost all the time.
 We present a theory and numerical evidence for
 this phenomenon.
 \end{abstract}
\pacs{05.45.-a, 05.45.Xt}
\maketitle
%
%

   Entrainment is
 a key mechanism for the emergence of order and coherence
 in a variety of physical systems
 consisting of oscillatory elements.
 It is one of the fundamental themes of nonlinear physics
 to explore the possible types of entrainments
 and clarify their fundamental properties.
 One of the typical entrainment phenomena
 is that caused by
 an external periodic signal.
 Consider two independent limit cycle oscillators
 of slightly different natural frequencies $\omega_i,~i=1,2$.
 Suppose that these two oscillators are in resonance with
 the external periodic signal, i.e.,
 $m\omega_i-n\Omega\simeq 0$,
 where $m$ and $n$ are integers and
 $\Omega$ is the external signal frequency.
 In this case,
 it is possible that
 the two oscillators have the same mean frequency $n\Omega/m$.
 This type of entrainment can be observed
 in various systems as diverse as
 periodically driven electrical circuits,
 lasers with coherent optical injections,
 and biological circadian rhythms.
 Resonant entrainment can be described
 by using simple dynamical models for phase variables
 and the essential properties are well understood
 (e.g., \cite{Kuramoto-1984}).

   Recent physical and numerical experiments
 have shown that
 not only a periodic but also a noise-like signal
 can give rise to
 entrainment between two independent oscillators
 \cite{Mainen-1995,Royama-1992,Yamamoto-2007,Teramae-2004,Nakao-2007}.
 The concept of entrainment of limit cycle oscillators
 induced by common signals
 has to be generalized to include
 the case of noise-like signals.
 The entrainment by a noise-like signal
 is a nonresonant one
 in the sense that 
 there is no resonance relation between
 the oscillator and the noise.
 Entrainment between
 two independent and {\it identical} oscillators
 induced by a common noise signal
 has already been studied \cite{Teramae-2004}:
 it has been analytically shown
 for a wide class of limit cycle oscillators that
 the phase locking state becomes linearly stable
 by applying an arbitrary weak Gaussian noise.

   However,
 in real systems,
 the two oscillators are never identical
 but slightly detuned.
 We note that
 the theory in Ref. \cite{Teramae-2004}
 does not guarantee
 the entrainment between nonidentical oscillators at all.
 As an illustrative example,
 consider two pairs of oscillators,
 where one pair has a natural frequency $\omega_1$
 and the other pair has
 a different natural frequency $\omega_2$,
 and suppose that
 they are subjected to a weak common noise.
 The theory
 tells that the phase locking occurs in each pair.
 However,
 it does not occur between the two pairs
 because for a weak noise
 the mean frequencies of
 the pair with $\omega_1$ and that with $\omega_2$
 are still close to $\omega_1$ and $\omega_2$, respectively.
 One might expect that
 the nonidentical oscillators come to
 have the same mean frequency and
 stable phase locking occurs when large enough noise is applied.
 As we will show, this is not the case.
 It has not yet been clarified at all
 what kind of phenomenon happens
 between nonidentical oscillators.
 It is necessary to clarify this point
 for better understanding of real systems.

   In this study,
 we consider a general class of limit cycle oscillators
 and reveal that
 a novel type of entrainment occurs
 between two nonidentical oscillators
 subjected to a common white Gaussian noise,
 which we call the {\it nonresonant entrainment}.
 This entrainment is anomalous in the sense that
 the two oscillators have different mean frequencies
 and the difference is constant
 even if the noise intensity increases
 but their phases come to be locked for almost all the time.
%
%
%

   Let $\mbox{\boldmath $X$}_i\in{\bf R}^N$ be
 a state variable vector and consider the equation
\begin{equation}
 \dot{\mbox{\boldmath $X$}_i}=
 \mbox{\boldmath $F$}(\mbox{\boldmath $X$}_i)
 +\delta\mbox{\boldmath $F$}_i(\mbox{\boldmath $X$}_i)
 +\mbox{\boldmath $G$}(\mbox{\boldmath $X$}_i)\eta(t),
 \quad i=1,2,
 \label{eqn:limit_cycle_Eq}
\end{equation}
 where
 $\mbox{\boldmath $F$}$ is an unperturbed vector field,
 $\delta\mbox{\boldmath $F$}_1$ and
 $\delta\mbox{\boldmath $F$}_2$ are small deviations from it,
 $\mbox{\boldmath $G$}\in{\bf R}^N$ is a vector function,
 and $\eta(t)$ is the white Gaussian noise
 such that
 $\langle \eta(t) \rangle = 0$ and
 $\langle \eta(t)\eta(s)\rangle = 2D\,\delta(t-s)$,
 where $\langle\cdots\rangle$ denotes
 averaging over the realizations of $\eta$
 and $\delta$ is Dirac's delta function.
 We call the constant $D>0$ the noise intensity.
 The noise-free unperturbed system
 $\dot{\mbox{\boldmath $X$}}
 =\mbox{\boldmath $F$}(\mbox{\boldmath $X$})$
 is assumed to have a limit cycle with a frequency $\omega$.
 We employ the Stratonovich interpretation for
 the stochastic differential equation (\ref{eqn:limit_cycle_Eq}).
 This interpretation allows us to apply
 the phase reduction method to Eq. (\ref{eqn:limit_cycle_Eq}),
 which assumes the conventional variable transformations
 in differential equations. 

   If we regard the common noise
 as a weak perturbation to the deterministic oscillators and
 apply the phase reduction method to Eq. (\ref{eqn:limit_cycle_Eq}),
 we obtain the equation for the phase variable as follows:
\begin{equation}
 \dot{\phi}_i=\omega+\delta\omega_i(\phi_i)+Z(\phi_i)\eta(t),
 \quad i=1,2,
 \label{eqn:phase_Eq_Str}
\end{equation}
 where $\omega$ is the frequency of the unperturbed oscillator,
 $\delta\omega_i$ is the frequency variation
 due to $\delta\mbox{\boldmath $F$}_i$,
 $Z$ is defined by
 $Z(\phi)=\mbox{\boldmath $G$}(\mbox{\boldmath $X$}_0(\phi))
 \cdot
 \left(
 \mbox{grad}_{\mbox{\tiny \boldmath $X$}}
 \phi|_{\mbox{\tiny\boldmath $X$}=\mbox{\tiny\boldmath $X$}_0(\phi)}
 \right)$,
 where $\phi$ is the phase variable defined
 by the unperturbed system
 $\dot{\mbox{\boldmath $X$}}=\mbox{\boldmath $F$}(\mbox{\boldmath $X$})$
 and $\mbox{\boldmath $X$}_0(\phi)$ is its limit cycle solution.
 By definition, $Z(\phi)$ is a periodic function,
 i.e., $Z(\phi)=Z(\phi+2\pi)$.
 We assume that
 $Z$ is three times continuously differentiable
 and not a constant.
 It is also assumed that $0<D/\omega\ll 1$
 to ensure the validity of the phase reduction.

   In order to derive the average equation for $\phi_i$,
 we translate Eq. (\ref{eqn:phase_Eq_Str}) into
 the equivalent Ito stochastic differential equation:
\begin{equation}
 \dot{\phi}_i=\omega+\delta\omega_i(\phi_i)
 +DZ(\phi_i)Z'(\phi_i)+Z(\phi_i)\eta(t),
 \label{eqn:phase_Eq_Ito}
\end{equation}
 where the dash denotes differentiation
 with respect to $\phi_i$.
 In the Ito equation, unlike in Stratonovich formulation,
 the correlation between $\phi_i$ and $\eta$ vanishes.
 If we subtract Eq. (\ref{eqn:phase_Eq_Ito}) for $\phi_2$
 from that for $\phi_1$
 and take the ensemble average,
 then we have the average equation
\begin{eqnarray}
 \!\!\!\!
 \frac{d}{dt}\langle\phi_1-\phi_2\rangle &=&
 \langle \delta\omega_1(\phi_1) \rangle
 - \langle \delta\omega_2(\phi_2) \rangle
\nonumber
\\
 &+&\!\!
 D\left\{
 \langle Z(\phi_1)Z'(\phi_1) \rangle
 -\langle Z(\phi_2)Z'(\phi_2) \rangle
 \right\},~
 \label{eqn:average_phase_Eq_Ito}
\end{eqnarray}
 where we used the fact $\langle Z(\phi_i)\eta(t) \rangle
 = \langle Z(\phi_i) \rangle\langle \eta(t) \rangle =0$.
 Each ensemble average on the right hand side
 can be evaluated by using
 the steady probability distribution
 $P_{i}(\phi_i)$ for $\phi_i$,
 which can be obtained from
 the Fokker-Planck equation for Eq. (\ref{eqn:phase_Eq_Ito}):
 i.e.,
 $\langle A(\phi_i) \rangle=\int_{0}^{2\pi}A(\phi)P_{i}(\phi)d\phi$,
 where $A$ represents a function of $\phi_i$.
 The distribution $P_{i}$ can be obtained as
 $P_{i}(\phi_i)=1/2\pi+O(\sigma_i,D/\omega)$,
 where
 $\sigma_i=\max_{0\le\phi<2\pi}|\delta\omega_i(\phi)/\omega|$.
 Since $\delta\mbox{\boldmath $F$}_i$ is small,
 $\sigma_i$ is a small parameter.
 Therefore,
 $P_{i}$ can be approximated by $P_{i}\simeq 1/2\pi$
 for small $D/\omega$
 and we have
\begin{eqnarray}
 \langle \delta\omega_i(\phi_i) \rangle
 &\simeq& \frac{1}{2\pi}\int_{0}^{2\pi}\delta\omega_i(\phi)d\phi
 \equiv \overline{\delta\omega}_i,
\label{eqn:average_delta_omega}
\\
 \langle Z(\phi_i)Z'(\phi_i) \rangle
 &\simeq& \frac{1}{2\pi}\int_{0}^{2\pi}Z(\phi)Z'(\phi)d\phi=0,
 \label{eqn:average_ZZ'}
\end{eqnarray}
 where we used the fact $Z(0)=Z(2\pi)$.
  If we substitute Eqs. (\ref{eqn:average_delta_omega})
 and (\ref{eqn:average_ZZ'}) into 
 Eq. (\ref{eqn:average_phase_Eq_Ito}), we have
\begin{equation}
 \frac{d}{dt}\langle\phi_1-\phi_2\rangle
 = \overline{\delta\omega}_1-\overline{\delta\omega}_2.
 \label{eqn:approx_average_phase_Eq_Ito}
\end{equation}
 Since in general
 $\overline{\delta\omega}_1-\overline{\delta\omega}_2\ne 0$,
 this equation indicates that
 the average phase difference
 increases or decreases in proportion to the time $t$.
 In other words,
 the two oscillators still have different mean frequencies
 even when a common white Gaussian noise is applied,
 i.e.,
 $d\langle\phi_1\rangle/dt \ne d\langle\phi_2\rangle/dt$.
 Intuitively, this result is natural
 because the white noise has a uniform power spectrum
 and does not have a characteristic frequency,
 which could entrain the oscillator frequencies.

   Let $\theta$ and $\psi$ be defined by
 $\theta=\phi_1-\phi_2$ and $\psi=\phi_1+\phi_2-2\omega t$.
 The variable $\theta$ measures the phase difference
 between the two oscillators.
 For small $D$ and $\overline{\delta\omega}_i$,
 it is expected that
 $\phi_i$ still has a mean frequency close to $\omega$.
 Therefore,
 $\theta$ and $\psi$ can be regarded as slow variables.
 If we change the independent variables
 form $(t,\phi_1,\phi_2)$ to $(t,\theta,\psi)$
 and perform the time-averaging with respect to $t$,
 we can obtain the Fokker-Planck equation
 corresponding to Eq. (\ref{eqn:phase_Eq_Str})
 as follows:
\begin{eqnarray}
 \frac{\partial Q}{\partial t}
 &=&
 -(\,\overline{\delta\omega}_1-\overline{\delta\omega}_2\,)
 \frac{\partial Q}{\partial \theta}
 -(\,\overline{\delta\omega}_1+\overline{\delta\omega}_2\,)
 \frac{\partial Q}{\partial \psi}
\nonumber
\\
 &&+D\frac{\partial^2}{\partial\theta^2}
 \bigl[\,2\{{\mit\Gamma}(0)-{\mit\Gamma}(\theta)\}Q\,\bigr]
\nonumber
\\
 &&+D\frac{\partial^2}{\partial\psi^2}
 \bigl[\,2\{{\mit\Gamma}(0)+{\mit\Gamma}(\theta)\}Q\,\bigr],
 \label{eqn:FP_Eq_theta}
\end{eqnarray}
 where $Q(t,\theta,\psi)$ is the joint probability distribution
 and ${\mit\Gamma}$ is defined by
\begin{equation}
 {\mit\Gamma}(\theta)=
 \frac{1}{2\pi}\int_{0}^{2\pi}Z(\phi)Z(\phi+\theta)d\phi.
 \label{eqn:def_Gamma}
\end{equation}
 Hereafter we assume the case of
 $\overline{\delta\omega}_1>\overline{\delta\omega}_2$
 without loss of generality.

   It is in general possible that
 $Z$ has a period smaller than $2\pi$.
 Since $Z$ is not a constant function,
 we suppose that $Z(\phi)=Z(\phi+2\pi/n)$,
 where $n$ is a positive integer. 
 Let $h(\theta)$ be defined by
 $h(\theta)=2\{{\mit\Gamma}(0)-{\mit\Gamma}(\theta)\}$.
 It can be shown that
 $h(\theta)\ge 0$ for any $\theta\in[0,2\pi)$.
 The zero points $s_m$ of $h$ are given by
 $s_m=2\pi m/n,~m=0,1,\dots,n-1$, where $s_0=0$.
 Equation (\ref{eqn:FP_Eq_theta}) has
 the steady solution $Q_s(\theta)$
 such that
 it is a continuous function of $\theta$ only
 and satisfies the two conditions
 ({\romannumeral 1}) $Q_s(\theta)=Q_s(\theta+2\pi)$ and
 ({\romannumeral 2}) $\int_{0}^{2\pi}Q_s(\theta)d\theta=1$.
 In each interval $(s_m,s_{m+1})$,
 the solution $Q_s$ can be obtained as follows:
\begin{equation}
 Q_s(\theta) =
 \frac{\varepsilon}{2\pi h(\theta)}\,
 \int_{\theta}^{s_{m+1}}\!\!
 \exp\biggl[\,-\varepsilon\int_{\theta}^{x}\frac{1}{h(y)}dy\,\biggr]dx,
 \label{eqn:steady_sol_Qs}
\end{equation}
 where
 $\varepsilon=(\overline{\delta\omega}_1-\overline{\delta\omega}_2)/D>0$.
 The right hand side of Eq. (\ref{eqn:steady_sol_Qs})
 has singularities at the zero points of $h$.
 The value of $Q_s$ for each $s_m$ is given by
 $Q_s(s_m)=\lim_{\theta\rightarrow s_{m}}Q_s(\theta)$.  
 Assume that $\theta\in(s_m,s_{m+1})$, i.e.,
 $\theta$ is an arbitrary regular point.
 It can be shown that
 $\lim_{\varepsilon\rightarrow 0}Q_s(\theta)=0$ holds
 due to the factor $\varepsilon$ in the numerator.
 This implies that
 the probability has to concentrates
 at the singular points $s_m,~m=0,1,\dots,n-1$
 because $Q_s$ satisfies the condition ({\romannumeral 2}).
 Thus,
 $Q_s$ in the limit $\varepsilon\rightarrow 0$
 is given by
\begin{eqnarray}
 Q_s(\theta)=\frac{1}{n}\sum_{m=0}^{n-1}\delta(\theta-s_m),
 \label{eqn:Qs_limit}
\end{eqnarray}
 where $\delta$ is Dirac's delta function.
 For small positive $\varepsilon$,
 the distribution $Q_s$ has narrow and sharp peaks
 at $\theta=s_m,~m=0,1,\dots,n-1$
 while $Q_s$ is close to zero
 in the regions other than
 the neighborhoods of these singular points.
 The peaks of $Q_s$ become narrower
 as $\varepsilon$ approaches zero.
 Equation (\ref{eqn:Qs_limit}) indicates that
 multiple peaks exist
 if $Z$ has a period smaller than $2\pi$, i.e., $n>1$.
 The existence of multiple peaks has been pointed out
 in the case of identical oscillators \cite{Nakao-2007}

   The above profile of $Q_s$ clearly shows that
 the phase locking states,
 where $\theta~\mbox{mod}~2\pi \simeq s_m$,
 are achieved for a large fraction of time
 during the time evolution
 when the noise intensity $D$ is relatively large
 with respect to the mean frequency difference
 $\overline{\delta\omega}_1-\overline{\delta\omega}_2$:
 i.e., the nonresonant entrainment occurs.
 Let $\delta$ be a small positive constant and
 $U_{\delta}$ be the $\delta$-neighborhood defined by
 $U_{\delta}=\cup_{m=0}^{n-1}(s_m-\delta,s_m+\delta)$,
 where  $\mbox{mod}~2\pi$ is taken for $s_0-\delta$.
 We identify the phase locking state
 by the condition $\theta\in U_{\delta}$.
 As shown by Eq. (\ref{eqn:approx_average_phase_Eq_Ito}),
 the present entrainment is not characterized by coincidence of
 the mean frequencies of the two oscillators.
 Therefore,
 as a measure for the entrainment,
 we introduce the phase locking time ratio $\mu$
 defined by
\begin{equation}
 \mu=\lim_{T\rightarrow \infty}\frac{T_L}{\,T\,},
 \label{eqn:def_mu}
\end{equation}
 where $T_L$ represents
 the total time length for which $\theta\in U_{\delta}$ happens
 during the period $T$.
 This ratio can also be expressed
 in terms of $Q_s$ by
 $\mu=\int_{U_{\delta}}Q_s(\theta)d\theta$,
 where the integral is taken over the set $U_{\delta}$.
 Equation (\ref{eqn:Qs_limit}) shows that $\mu\rightarrow 1$
 in the limit $\varepsilon=
 (\overline{\delta\omega}_1-\overline{\delta\omega}_2)/D
 \rightarrow 0$.

   A phase locking state
 cannot continue for the infinite time
 but phase slips have to happen during the periods
 such that $\theta\notin U_{\delta}$
 because the two mean frequencies
 $d\langle\phi_1\rangle/dt$ and $d\langle\phi_2\rangle/dt$
 are different.
 Equation (\ref{eqn:approx_average_phase_Eq_Ito})
 indicates that the mean frequency difference is
 given by the constant
 $\overline{\delta\omega}_1-\overline{\delta\omega}_2$.
 This implies that
 the average number of phase slips,
 which happen in a unit time interval,
 does not become small but remains constant
 even for relatively large $D$
 compared with
 $\overline{\delta\omega}_1-\overline{\delta\omega}_2$.
 In other words,
 the average interslip interval remains constant.
 On the other hand,
 the probability for $\theta\notin U_{\delta}$
 decreases and converges to zero as $D$ increases:
 i.e., the phases come to be locked for almost all the time.
 These two facts imply that
 a single phase slip completes more rapidly:
 i.e.,
 the time needed for one phase slip decreases and converges to zero
 as $D$ increases.
 We emphasize that
 the above mentioned behavior is
 a remarkable feature of the nonresonant entrainment.
 This behavior is very different from
 that of resonant entrainment by a periodic signal,
 where the average interslip interval diverges
 and the mean frequencies becomes identical
 as the signal intensity approaches
 the critical value for entrainment.
%
%
%

   In order to demonstrate the above analytical results,
 we show numerical results
 for an example described
 by the Stratonovich stochastic differential equations
\begin{equation}
 \dot{\phi}_i=\omega_i+\sin(\phi_i)\,\eta(t),
 \quad i=1,2,
 \label{eqn:phase_Eq_sin}
\end{equation}
 where $\omega_i,~i=1,2$ are slightly different constants.
%
%
\begin{figure}[t]
 \includegraphics[width=80mm,height=50mm]{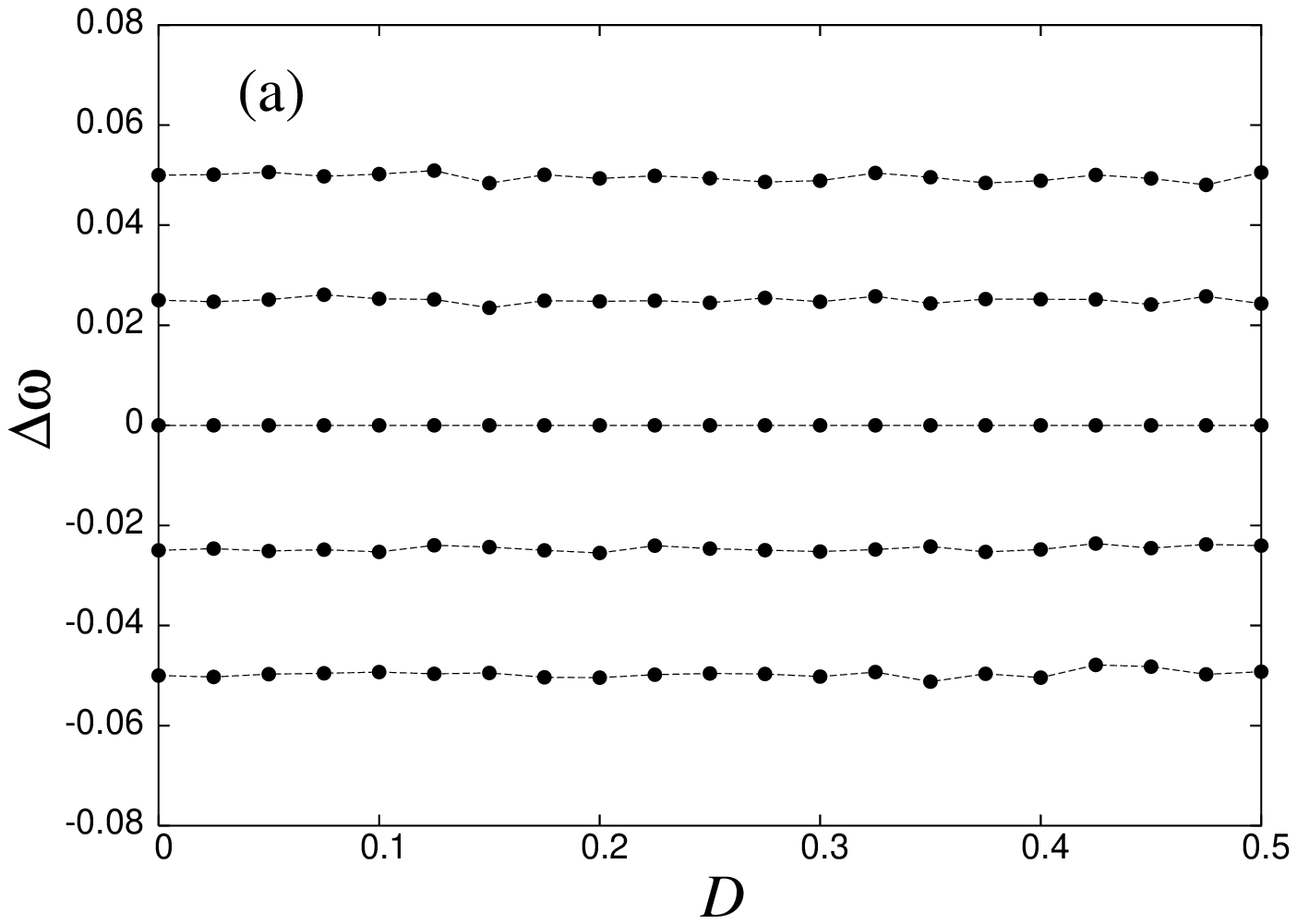}
 \includegraphics[width=80mm,height=50mm]{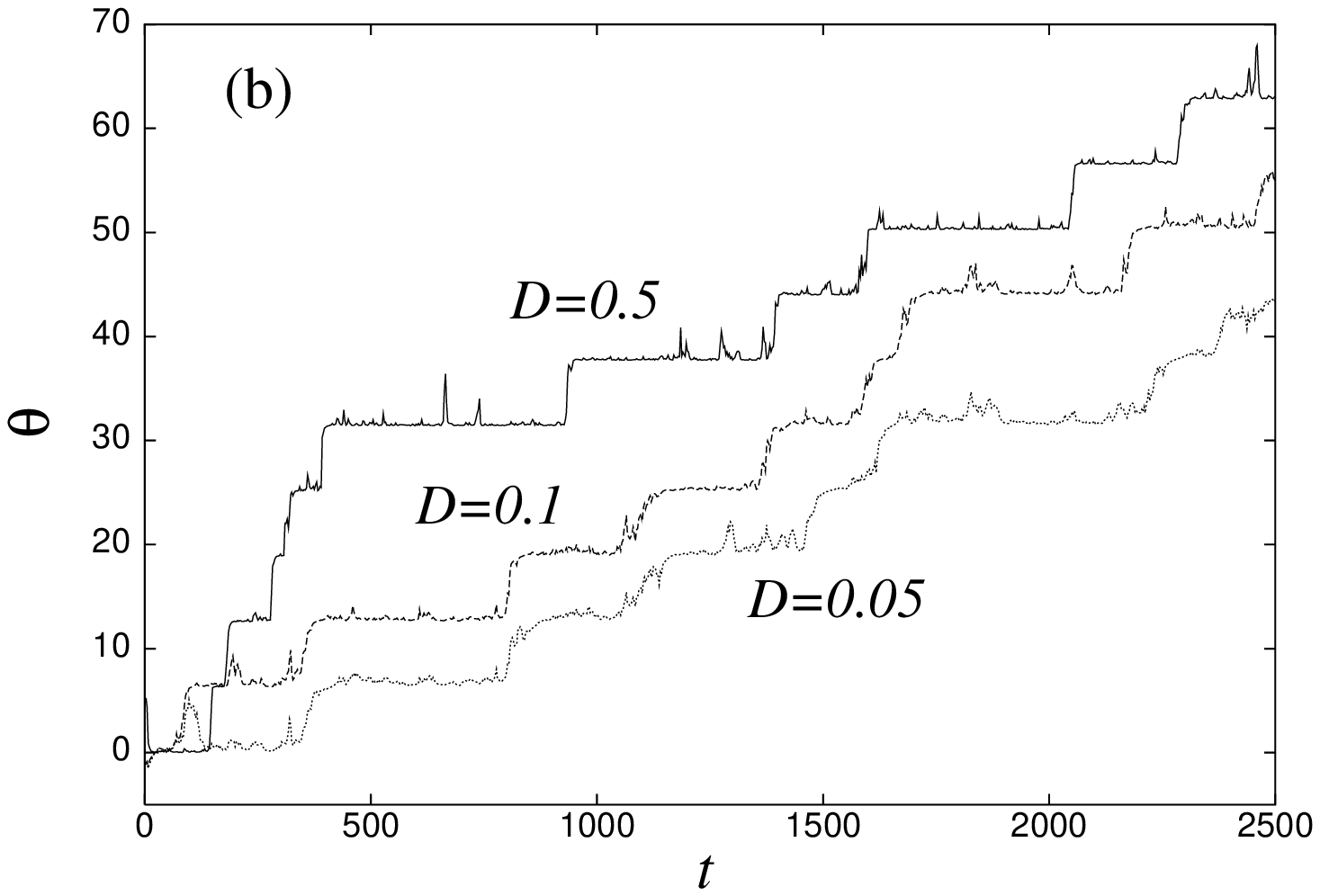}
 \includegraphics[width=80mm,height=55mm]{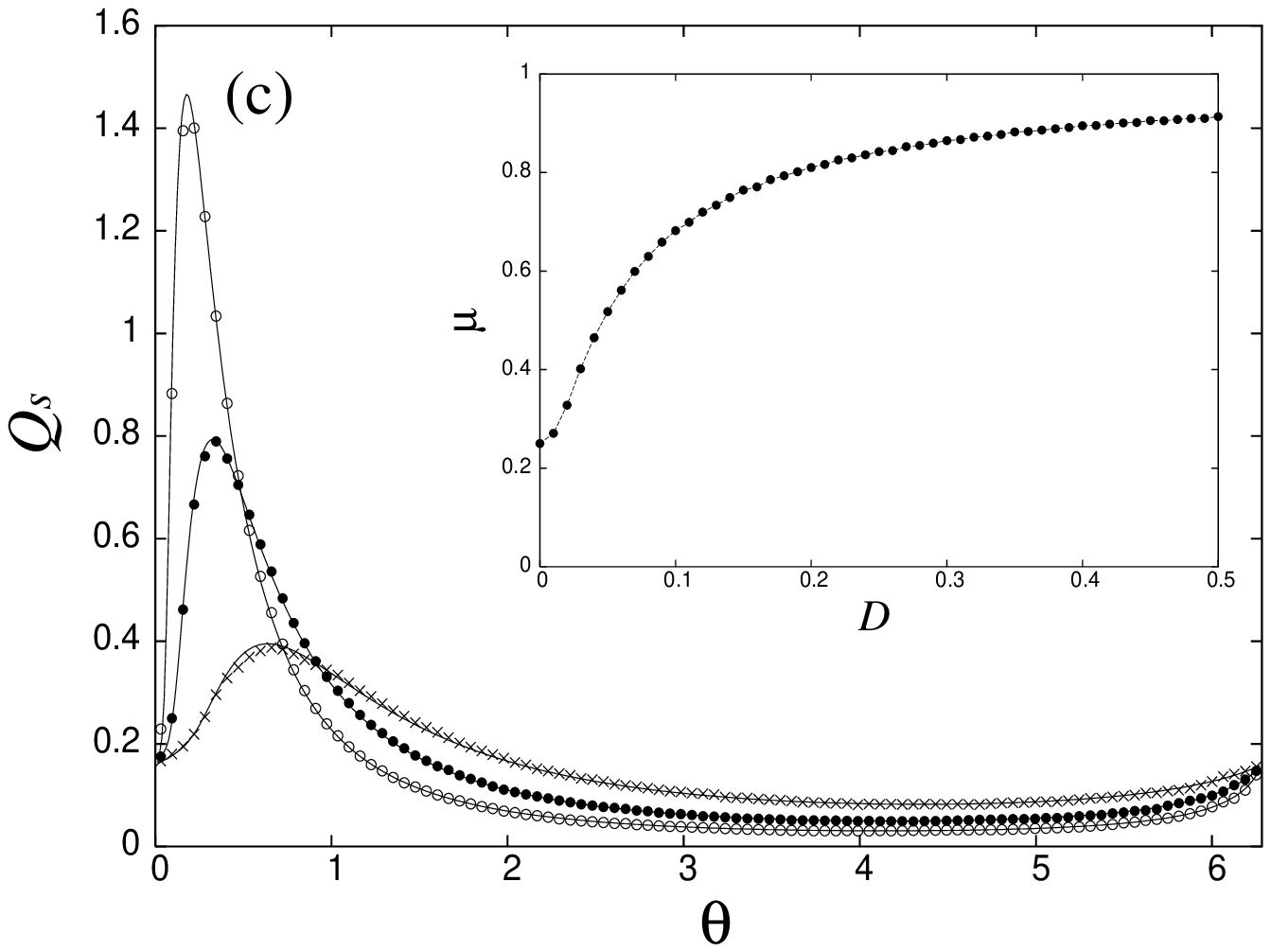}
 \caption{Entrainment in phase models with $Z=\sin(\phi)$:
 (a) mean frequency difference $\Delta\omega$ vs. $D$,
 (b) time evolution of phase difference $\theta=\phi_1-\phi_2$
 and
 (c) probability distribution $Q_s(\theta)$ for
 $D=0.02\,(\times),~0.05\,(\bullet),~\mbox{and}~0.1\,(\circ)$,
 where analytical results are shown by solid line.
 The inset in (c) shows $\mu$ plotted against $D$.
 In (b) and (c), $\omega_1=1$ and $\omega_2=0.98$.
 }
 \label{fig:example_phase_osc}
\end{figure}

   Figure \ref{fig:example_phase_osc}(a) shows
 the mean frequency difference
 $\Delta\omega=d\langle\phi_1\rangle/dt-d\langle\phi_2\rangle/dt$
 plotted as a function of $D$,
 where $\omega_1$ is fixed to unity
 and five different values of $\omega_2$ are employed.
 We calculated $\Delta\omega$
 by replacing the ensemble averages in $\Delta\omega$
 with the time averages,
 i.e., $d\langle\phi_i\rangle/dt=
 \lim_{T\rightarrow\infty}\{\phi_i(T)-\phi_i(0)\}/T$.
 The mean frequency difference $\Delta\omega$
 is not zero except for the case $\omega_1=\omega_2=1$.
 It is clearly shown that
 $\Delta\omega$ is constant and independent of $D$.
 This result coincides with the analytical result of
 Eq. (\ref{eqn:approx_average_phase_Eq_Ito}) .
 The steady distribution $P_i(\phi_i)$ is
 approximately given by
 $P_i(\phi_i)\simeq(1/2\pi)[\,1+(D/2\omega_i)\sin(2\phi_i)\,]$
 for this example.
 This shows that
 the assumption $P_i(\phi_i)\simeq 1/2\pi$
 is reasonable for small $D$ used in the numerical calculations.
 Thus, the result of
 Eq. (\ref{eqn:approx_average_phase_Eq_Ito}) holds.
 
   The time evolution of
 the phase difference $\theta=\phi_1-\phi_2$ 
 is shown for three different values of $D$
 in Fig. \ref{fig:example_phase_osc}(b),
 where $\omega_1=1$ and $\omega_2=0.98$.
 These results clearly show that
 the phases are locked near $\theta\simeq 2\pi n,~n\in{\bf Z}$
 and the phase slips occur intermittently. 
 It should be noted that
 the time needed for a single phase slip
 becomes smaller as $D$ increases.
 This observation is in agreement
 with the analytical result.

   The probability distribution $Q_s(\theta)$
 is shown in Fig. \ref{fig:example_phase_osc}(c)
 for three different values of $D$,
 where $\omega_1=1$ and $\omega_2=0.98$.
 The analytical results of Eq. (\ref{eqn:steady_sol_Qs}) 
 are also shown for the corresponding values of 
 $\varepsilon=(\omega_1-\omega_2)/D$.
 It is seen that
 $Q_s$ is close to the uniform distribution
 for small $D$ or large $\varepsilon$.
 In contrast,
 the distribution has a sharp peak near $\theta=0$ 
 for large $D$ or small $\varepsilon$.
 The peak in $Q_s$ becomes narrower
 and its position becomes closer to $\theta=0$
 as $D$ increases.
 This agrees with the previous theory
 since the zero point of $h(\theta)$ is only $\theta=0$
 in this example
 and thus the theory tells that
 $Q_s$ has a peak only at $\theta=0$.
 It is also seen that
 the peak is not centered at $\theta=0$
 but shifted to the positive direction:
 i.e.,
 the phase $\phi_1$ of the larger natural frequency oscillator
 is kept advanced with respect to $\phi_2$
 even in the phase locking state. 
 The inset of Fig. \ref{fig:example_phase_osc}(c)
 shows that
 the phase locking time ratio $\mu$
 monotonically increases and approaches unity
 with increasing $D$.
 Figure \ref{fig:example_phase_osc}(c) clearly demonstrates
 that the phases are locked
 for a larger fraction of the time
 as $D$ increases.
%
%
%
%
\begin{figure}[t]
 \includegraphics[width=80mm,height=55mm]{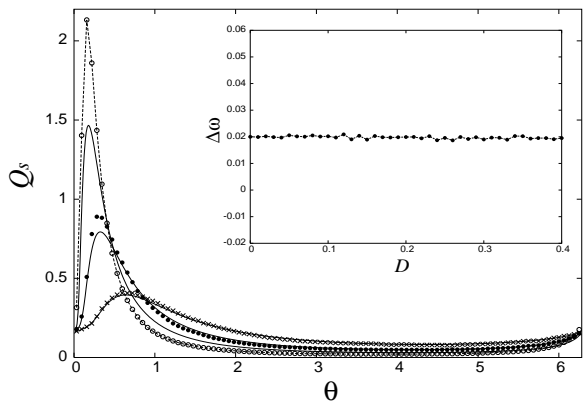}
 \caption{Entrainment in SL oscillators.
 Probability distribution $Q_s(\theta)$ is shown for
 $D=0.02\,(\times),~0.05\,(\bullet),~\mbox{and}~0.1\,(\circ)$,
 where analytical results are shown by solid line.
 The inset shows $\Delta\omega$ vs. $D$.
 Parameters are $\delta\omega_1=0$ and $\delta\omega_2=-0.02$.
 }
 \label{fig:example_SL_osc}
\end{figure}

  In order to validate
 the theory based on the phase reduction method,
 we carried out numerical experiments
 for the Stuart-Landau (SL) oscillator
\begin{equation}
 \dot{\psi}_j =
 (1+ic_j)\psi_j-|\psi_j|^2\psi_j-\eta(t),
 \quad j=1,2,
 \label{eqn:Complex_Stuart-Landau_osc}
\end{equation}
 where $\psi_j\in{\bf C}$ and
 $c_j=1+\delta\omega_j$ is a real constant.
 This is reduced to the phase model
 $\dot{\phi}_j=1+\delta\omega_j+\sin(\phi_j)\eta(t)$,
 where $\phi_j$ is the appropriately defined phase variable.

   In Fig. \ref{fig:example_SL_osc},
 the numerically obtained distribution $Q_s(\theta)$
 is shown for three different values of $D$,
 where $\delta\omega_1=0$ and $\delta\omega_2=-0.02$.
 The analytical results
 obtained from the corresponding phase model
 are also shown for the corresponding values of 
 $\varepsilon=(\delta\omega_1-\delta\omega_2)/D$.
 A sharp peak of $Q_s$ appears near $\theta=0$.
 It becomes narrower and approaches $\theta=0$
 as $D$ increases.
 Agreement between the numerical and analytical results
 is excellent, especially in small $D$ region,
 where the phase reduction method gives a good approximation.
 The inset shows the mean frequency difference
 $\Delta\omega=d\langle\phi_1\rangle/dt-d\langle\phi_2\rangle/dt$
 plotted as a function of $D$
 for the same $\delta\omega_1$ and $\delta\omega_2$. 
 It is clearly shown that
 $\Delta\omega$ does not depend on $D$ and
 its constant value is given by
 $\Delta\omega=\delta\omega_1-\delta\omega_2$.
 This behavior also agrees with the theory.
 The agreements in the behaviors of $Q_s$ and $\Delta\omega$
 validate the theory based on the phase model.
%
%
%

   In conclusion,
 we have found the nonresonant entrainment
 between two nonidentical limit cycle oscillators
 subjected to a common external white Gaussian noise.
 We theoretically elucidated this phenomenon
 by using a phase model and
 presented numerical evidence
 for a particular phase model and the SL oscillator.
 The nonresonant entrainment
 is anomalous in the sense that
 the two oscillators have different mean frequencies,
 where the difference is independent of the noise intensity
 and the average interslip interval is constant,
 while their phases come to be locked for almost all the time
 for relatively large noise.
 It is expected that
 a similar entrainment occurs
 for various noise-like signals
 having broad continuous power spectra.
%
%
%
%

   The authors would like to thank R. Roy,
 D. Tanaka, H. Nakao, and T. Aoyagi for helpful discussions.
%
%
%
%

%
%
\end{document}